\documentclass[sigconf]{acmart}
\usepackage[utf8]{inputenc}
\usepackage{hyperref}
\usepackage{url}
\usepackage[neverdecrease]{paralist}
\usepackage[caption=false]{subfig}
\usepackage{graphicx}
\usepackage{booktabs}
\usepackage{graphics}
\usepackage{tabularx}
\usepackage{xspace} 
\usepackage{graphicx}
\usepackage{graphicx} 
\usepackage{booktabs}
\usepackage{mdwlist}
\usepackage{multirow}
\usepackage{amsmath}
\usepackage{enumitem} 
\usepackage{calc}
\usepackage{colortbl}
\usepackage[ruled,linesnumbered]{algorithm2e}
\usepackage{booktabs}
\usepackage{mathtools}
\usepackage{pifont}
\usepackage{color,soul}
\usepackage{array}
\usepackage{adjustbox}
\usepackage{rotating}
\usepackage{colortbl}

\usepackage[show]{chato-notes}

\newcommand{\para}[1]{\paragraph{\textnormal{\textbf{#1}}}} 
\newcommand\defas{\mathrel{\overset{\makebox[0pt]{\mbox{\normalfont\tiny\sffamily def}}}{=}}}

\usepackage[neverdecrease]{paralist}
\newcommand{\uls}{\begin{itemize}[leftmargin=*]}
\newcommand{\ule}{\end{itemize}}
\newcommand{\ols}{\begin{enumerate}[leftmargin=*]}
\newcommand{\ole}{\end{enumerate}}
\newcommand{\li}{\item}

\newcommand\MyBox[2]{
  \fbox{\lower0.75cm
    \vbox to 1.7cm{\vfil
      \hbox to 1.7cm{\hfil\parbox{1.4cm}{#1\\#2}\hfil}
      \vfil}%
  }%
}

\copyrightyear{2023}
\acmYear{2023}
\setcopyright{acmlicensed}\acmConference[SIGIR '23]{Proceedings of the 46th International ACM SIGIR Conference on Research and Development in Information Retrieval}{July 23--27, 2023}{Taipei, Taiwan}
\acmBooktitle{Proceedings of the 46th International ACM SIGIR Conference on Research and Development in Information Retrieval (SIGIR '23), July 23--27, 2023, Taipei, Taiwan}
\acmPrice{15.00}
\acmDOI{10.1145/3539618.3592046}
\acmISBN{978-1-4503-9408-6/23/07}


\title{Query-specific Variable Depth Pooling via Query Performance Prediction towards Reducing Relevance Assessment Effort}

\author{Debasis Ganguly}
\affiliation{%
  \institution{University of Glasgow}
  \city{Glasgow}
  \country{United Kingdom}
}
\email{debasis.ganguly@glasgow.ac.uk}

\author{Emine Yilmaz}
\affiliation{%
  \institution{University College London}
  \city{London}
  \country{United Kingdom}
}
\email{emine.yilmaz@ucl.ac.uk}

\begin{document}

\begin{abstract}
Due to the massive size of test collections, a standard practice in IR evaluation is to construct a `pool' of candidate relevant documents comprised of the top-$k$ documents retrieved by a wide range of different retrieval systems - a process called depth-$k$ pooling. A standard practice is to set the depth ($k$) to a constant value for each query constituting the benchmark set. However, in this paper we argue that the annotation effort can be substantially reduced if the depth of the pool is made a variable quantity for each query, the rationale being that the number of documents relevant to the information need can widely vary across queries.
Our hypothesis is that a lower depth for the former class of queries and a higher depth for the latter can potentially reduce the annotation effort without a significant change in retrieval effectiveness evaluation. We make use of standard query performance prediction (QPP) techniques to estimate the number of potentially relevant documents for each query, which is then used to determine the depth of the pool. Our experiments conducted on standard test collections demonstrate that this proposed method of employing query-specific variable depths is able to adequately reflect the relative effectiveness of IR systems with a substantially smaller annotation effort.  

\end{abstract}



\maketitle

\section{Introduction}

The most widely used approach used in evaluating quality of retrieval systems is based on constructing test collections via the Cranfield paradigm \cite{Cleverdon-The-1967}, which assumes that relevance judgments for each query are complete. However, due to the cost of obtaining relevance judgments, it is often impractical to obtain relevance judgments for all documents in a collection. 

A commonly used approach to reduce the need for extensive judging effort in test collection construction is depth-$k$ pooling, which is based on constructing a pool of documents that consists of the top-$k$ documents retrieved by various systems and then obtaining relevance judgments for these documents, assuming that the rest of the documents are non-relevant. Most existing test collections, such as the ones constructed by TREC are constructed using depth-100 pools, i.e., using a depth of $k=100$   \cite{TREC8_Overview}. However, depth-100 pools still tend to be quite large and hence significant research has been devoted to reducing the number of judgments needed in constructing test collections \cite{YilmazKA08, CarteretteAS06}. In order to reduce the pool depth, some recent test collections are instead constructed using much shallower depths, such as depth-10 pools used by the recent Deep Learning Track collections \cite{craswell2020overview}.

Most previous work assumes that a constant depth ($k$) should be used across all the queries in the test collection. However, some queries may contain more relevant documents than the others, and using the same depth across all queries could lead to wasting a significant proportion of annotation budget on those queries where fewer judgments could have been sufficient \cite{YilmazKA08,Yilmaz14}. 

Some previous work based on active learning proposed approaches employed deeper depths for systems that are more likely to retrieve a higher number of relevant documents \cite{Sanderson10,Losada16,Cormack98} - a process which often leads to \emph{different rank cutoffs for different systems}.
Optimising the resultlist presentation to search system users motivated a
similar thread of work that involves chopping off a ranked list of documents at different cut-off points based on the statistics of their score distributions \cite{stop_reading,truncation_rl}. More recently, supervised learning via neural networks has been applied to address this problem \cite{choppy}.

While previous work has investigated cutting off ranked lists at variable depths to reduce the information finding effort of search engine users \cite{Sanderson10, Losada16, Cormack98, stop_reading, choppy}, we, in contrast, use the concept of variable depths for reducing the assessment effort. The rationale behind the idea is that some queries have a smaller number of documents in the collection that are relevant to their corresponding information needs, whereas for some other queries this number may be substantially larger. Our hypothesis is that a lower depth for the former class of queries and a higher depth for the latter can potentially reduce the annotation effort without a significant change in the relative evaluation of different retrieval systems.

In particular, to estimate the number of potentially relevant documents for each query we make use of a standard query performance prediction (QPP) based approach (specifically, NQC \cite{kurland_tois12} in this paper). This number of relevant documents estimated for a query in the top-documents retrieved by an IR system is then used to determine the number of documents that contribute to the pool for that particular query and IR system combination.




\section{Proposed Methodology}


\subsection{A Review of QPP}

QPP approaches can broadly be categorized into the pre-retrieval and post-retrieval types.
A pre-retrieval estimator uses aggregated collection-level statistics (e.g., maximum or average of the inverse document frequencies of the query terms) as an estimated performance measure of a query \cite{pre-ret_survey_cikm08,qpp_preret_springer,qpp_preret_ecir08}.
A post-retrieval estimator, on the other hand, makes use of the information from the set of top-retrieved documents to estimate the quality of the retrieved list. In general, the QPP score for a post-retrieval estimator $\phi$ is a function of the query and the set of top-retrieved documents, i.e.,
\begin{equation}
\phi: Q \times M^{(k)}(Q) \mapsto \mathbb{R} \label{eq:generic},
\end{equation}
where $M^{(k)}(Q)$ denotes the set of top-$k$ documents retrieved for query $Q$ with a model $M$. From hereon, $M^{(k)}(Q)$ is abbreviated as $M^{(k)}$, the query being understood from the context.

Various evidences extracted from the top-retrieved documents have been shown to be useful for different post-retrieval QPP estimation methods, such as the KL divergence between the language model of the top-retrieved documents and the collection model as in Clarity \cite{croft_qpp_sigir02}, the
aggregated values of the information gains of each top-retrieved document with respect to the collection as in WIG (Weighted Information Gain) \cite{wig_croft_SIGIR07}, the skewness of the retrieval status values (RSVs) measured with variance as in NQC (Normalized Query Commitment) \cite{kurland_tois12}, ideas based on the clustering hypothesis for a pairwise document similarity matrix \cite{fernando_correlation_sigir07}, topology of the embedded word vectors \cite{RoyGMJ16} and even supervised approaches using neural networks \cite{datta_deepqpp22,DBLP:conf/sigir/DattaMGG22}.

In our work, as an unsupervised QPP approach we employ NQC (Normalized Query Commitment) \cite{kurland_tois12}, which is a simple yet effective post-retrieval QPP method \cite{query_variants_kurland,query_variants_suchana} (we leave the investigation with other QPP approaches as future work).
NQC predicts the retrieval effectiveness of a query using the variance of the document scores, the rationale being that a query with a well-defined information need is likely to lead to a more non-uniform (heavy-tailed) distribution of the RSVs.
Formally speaking, the generic $\phi$ function of Equation \ref{eq:generic} takes the form
\begin{equation}
\phi_{\mathrm{NQC}}(Q, M^{(k)}) \defas
\frac{\sqrt{
\frac{1}{k}\sum_{i=1}^k (P(D_i|Q)-\bar{P}(D|Q))^2}
}
{
P(Q|C)
},
\label{eq:nqc}
\end{equation}
where
$P(D_i|Q)$ denotes the RSV of the document $D_i$ to $Q$, $\bar{P}(D|Q)$ denotes the mean of the RSVs, and $P(Q|C)$ denotes the similarity of $Q$ to the collection, which is computed by aggregating collection statistics (e.g., idf) over the query terms.

Although for our experiments we specifically use the NQC method, our proposed method of
variable depth pooling strategy (to be discussed in the next section)
is a general one allowing application of any other QPP model as a concrete realisation of $\phi$ (Equation \ref{eq:nqc}).


\subsection{Depth Estimation using QPP}

Given a set of $n$ queries $\mathcal{Q} = \{Q_1,\ldots,Q_n\}$, a standard depth-$k$ pooling process first involves employing a number of different retrieval systems (models), say $\mathcal{M}=\{M_1,\ldots,M_p\}$ to construct a \emph{pool} of the top-$k$ documents retrieved with each $M_i$ ($i=1,\ldots,p$). Formally speaking, this pool of depth $k$, $\mathcal{P}_k(Q)$ for query $Q \in \mathcal{Q}$ is constructed as
\begin{equation}
\mathcal{P}_k(Q) = \cup_{i=1}^p M^{(k)}_i,   \label{eq:pool}
\end{equation}
where $M^{(k)}_i$ denotes the top-$k$ documents retrieved with model $M_i$.

The key idea now is to make this depth $k$ a function of $Q$ itself rather than it being a constant across all queries. We propose to make this integer-valued depth of a query a function of the generic form of the real-valued QPP estimate $\phi$ (which depends on $M^{(k)}_i$) as shown in Equation \ref{eq:generic}, and denote this integer-valued depth function as $\zeta(Q, M_i) \in \mathbb{Z}^+$.

One important point to note is that we make the variable depth a function of the query and of the retrieval model.
Specifically, after computing the depth $\zeta(Q, M_i)$ for a `query and system' combination $(Q, M_i)$, we use this depth to determine the number of documents top documents from $M^{(\zeta(Q, M_i))}_i$ to add to the pool. In other words, we obtain a more generic version of Equation \ref{eq:pool} as
\begin{equation}
\mathcal{P}_{\zeta(Q)} = \cup_{i=1}^p M^{(\zeta(Q, M_i))}_i.
\end{equation}

We now explore two different ways by which this variable depth of a query may depend on the QPP estimator. Each of these two choices of $\zeta(Q, M)$ has its own set of supporting arguments; more details follow.

\subsubsection{Inverse Linear Dependence}
The first choice for $\zeta(Q, M)$ is a linearly inverse proportional function, the intuition for which is that the higher the value of the estimate - the higher is the likelihood of the ranked list (as retrieved by $M$) to contain a higher proportion of relevant documents towards the top ranks. This, in turn, means that a smaller depth for such a query is likely to be adequate to include an adequate set of potentially relevant documents in the constructed pool for a robust evaluation of IR systems.

On the other hand, a relatively low value of the QPP estimate for a query potentially indicates that more documents should perhaps be included in the pool by employing a higher depth value for that query.
Formally speaking, using the generic notation of the QPP function of Equation \ref{eq:generic}, the depth of a query $Q$ is then
\begin{equation}
\zeta(Q, M_i) = d_{min} + \lfloor (1-\phi(Q, M^{(d_{max})}_i)) (d_{max} - d_{min})\rfloor, \label{eq:vardepth}    
\end{equation}
where the parameters $d_{min}$ denotes the minimum depth, and $d_{max}$ denotes the maximum depth ($d_{min}, d_{max} \in \mathbb{Z}^+$, i.e., they both are positive integers).

Normalized values of the QPP $\phi(Q, M^{(d_{max})}_i)$ estimates ensure that the depth of a query is an integer between the integer bounds $d_{min}$ and $d_{max}$. Note that, in particular, for computing the QPP estimates themselves we use $k=d_{max}$ (the maximum depth), and we apply max-normalization for the QPP estimates. 

\subsubsection{Linear Dependence}
The argument for this choice of $\zeta(Q, M)$ is that a higher value of $\phi(Q, M^{(k)})$ is likely to indicate that a higher number of potentially relevant documents for $Q$ exists in the collection. This, in turn, means that one may consider probing at higher depths to collect those candidates for assessment for a more comprehensive evaluation of IR systems. Similarly, a smaller estimate for $\phi(Q, M^{(d_{max})})$ means that it is not worthwhile to use a high depth for $Q$ because the candidates collected from lower down the ranked lists of such IR systems may end up in the ground-truth set indicating wasted manual effort. Formally,
\begin{equation}
\zeta(Q, M_i) = d_{min} + \lfloor \phi(Q, M^{(d_{max})}_i) (d_{max} - d_{min})\rfloor, \label{eq:vardepth_direct}    
\end{equation}
where the only difference of Equation \ref{eq:vardepth_direct} with that of \ref{eq:vardepth} is that in the former an increase in the $\phi(Q, M^{(d_{max})}_i)$ increases the depth to a higher integer value within the bounds $[d_{min}, d_{max}]$ instead of decreasing it as is the case for the latter.

\section{Evaluation} \label{sec:eval}

\subsection{Experiment Details}
\subsubsection{Research Questions}
We conduct experiments to investigate the following two research questions.
\uls
\li \textbf{RQ-1}: Is an NQC-based variable depth pooling strategy beneficial to reduce annotation effort without causing significant changes in the relative system ranks?
\li \textbf{RQ-2}: Which depth selection function (linear or inverse linear - Equations \ref{eq:vardepth} or \ref{eq:vardepth_direct}) turns out to be the more effective of the two?   
\ule

\subsubsection{Datasets}
Our retrieval experiments are conducted on two standard datasets used for the ad-hoc IR task, namely the TREC Robust \cite{TREC8_Overview} and the TREC DL datasets \cite{msmarco-data}.
While the former is comprised of news articles, the latter is a collection of passages accumulated with Bing queries.
The set of relevant documents comprising the ground-truth of the TREC Robust dataset was constructed via depth-100 pooling \cite{TREC8_Overview}. On the other hand, in TREC DL a combination of depth-10 pooling and an active learning based strategy \cite{activelearning} was used to compile the ground-truth \cite{trecdl2019,trecdl2020}. We leave out TREC 6 topic sets from our experiments for consistency with the remaining topic sets, the ground-truths of which do not include the congressional records (CR).
Table \ref{tab:dataset} summarises the datasets used for our experiments.

\subsubsection{Setup}
For each topic set used in our experiments, we make use of the officially submitted runs as downloaded from the TREC archive\footnote{\url{https://trec.nist.gov/results/}}. We conduct our experiments on each topic set separately so as to compute the effect of the relative changes in the systems (officially submitted runs) in each.

In each experiment, the value of the minimum depth ($d_{min}$ of Equations \ref{eq:vardepth} and \ref{eq:vardepth_direct}) was set to $10\%$ of the true depth used to construct the pool of the respective datasets, i.e., $0.1\times100=10$ for TREC Robust, and $0.1\times10=1$ for TREC DL.
Similarly, the value of the maximum depth ($d_{max}$) was set to half the value of the true depths, i.e., $100/2=50$ and $10/2=5$ for TREC Robust and TREC DL datasets, respectively.
As the QPP estimate $\phi(Q, M^{(k)})$, we use the standard unsupervised QPP approach - NQC.

\begin{table}[t]
\centering
\caption{
\small
Summary of the datasets used in our experiments. The columns `$\bar{|Q|}$' and `$\bar{\#Rel}$' denote the average number of query terms and average number of relevant documents, respectively. The column $p$ denotes the number of official runs submitted, all of which is used to construct the pools (Equation \ref{eq:pool}) for each topic set.}
\begin{adjustbox}{width=0.9\columnwidth}
\small
\begin{tabular}{@{}lclcccc@{}}
\toprule
Collection & \#Docs & Topics & \#Topics & $\bar{|Q|}$ & $\bar{\#Rel}$ & $p$\\

\midrule

Robust & \multirow{2}{*}{528,155} & TREC 7 & 50 & 2.42 & 93.48 & 103\\
(disks 4,5 - CR) & & TREC 8 & 50 & 2.38 & 94.56 & 129\\
\midrule

MS MARCO & \multirow{2}{*}{8,841,823} & DL'19 & 43 & 5.40 & 58.16 & 37 \\
Passage & & DL'20 & 54 & 6.04 & 30.85 & 59\\
\bottomrule
\end{tabular}

\label{tab:dataset}
\end{adjustbox}
\end{table}


\subsubsection{Pooling Methods Investigated}
%
As baselines, we employ the standard procedure of constant-depth pooling (\textbf{CDP}) (Equation \ref{eq:pool}). Since our proposed methodology uses depths that varies across queries, for a fair comparison we compare our proposed approach with several CDP baselines, as enumerated below.
\uls
\li \textbf{CDP-Max} involves setting $k=d_{max}$ in Equation \ref{eq:pool}, where $d_{max}$ is the upper bound of the depth used in the VDP approach ($d_{max}$ in Equations \ref{eq:vardepth} and \ref{eq:vardepth_direct}). This method thus represents an apex-line or \textit{oracle} scenario with a larger pool size thus implying a larger effort for relevance assessments.
\li \textbf{CDP-Min} is a baseline which sets $k=d_{min}$ in Equation \ref{eq:pool} thus implying that this represents the lower end of the spectrum with a much smaller pool size.
\li \textbf{CDP-Avg} is a baseline with the depth of the pool being set to the closest integer of the mid-point of the interval $[d_{min}, d_{max}]$, i.e., setting $k=\lfloor (d_{max}-d_{min})/2 \rfloor$ in Equation \ref{eq:pool}. This baseline yields a pool that is expected to be of a size similar to those obtained by the VDP-based methods.
\ule
%
As variants of our proposed methodology of variable-depth pooling (\textbf{VDP}), we explore the following.
\uls
\li \textbf{VDP-IL}: this denotes variable-depth pooling by means of an inverse linear dependence (Equation \ref{eq:vardepth}).
\li \textbf{VDP-L}: this denotes variable-depth pooling with a linear dependence (Equation \ref{eq:vardepth_direct})\footnote{The implementation of all the methods investigated is available at \url{https://github.com/gdebasis/vardepthpooling}}.
\ule

\subsubsection{Evaluation Metrics}
As per the standard practice of a simulated pooling setup \cite{DBLP:conf/cikm/AslamY07}, the pool of documents obtained with each method is a subset of the existing relevance assessments. This allows provision to compute the quality of a pool by comparing the correlation of the relative system ranks measured via the ground-truth induced on the subset as against the entire existing pool of the respective datasets.

A smaller pool is considered to be of good quality if the relative system ranks measured via an IR metric (e.g., AP) on this smaller set of ground-truth does not change substantially in comparison to those measured with the larger pool. In particular, as correlation measures between IR models we employ Pearson's $r$ and Kendall's $\tau$. We employed mean average precision (MAP) to induce an order on the different officially submitted runs (systems). As per the standard practice, AP on the TREC DL dataset treated graded judgments higher than or equal to 2 as relevant \cite{trecdl2019,trecdl2020}. 

In addition to the relative rank stability of systems, we also report the recall or \textbf{coverage}, measured as the fraction of relevant documents found in a depth restricted pool averaged across all the queries of a benchmark topic set.
Formally,
\begin{equation}
\mathcal{C}=
\frac{1}{|R_{max}(\mathcal{Q})|}
\sum_{Q \in \mathcal{Q}}
\sum_{D \in \mathcal{P}_\zeta(Q)}\mathbb{I}(\mathrm{Rel}(D,Q)=1)
\label{eq:recall},
\end{equation}
where $\mathrm{Rel}(D,Q)=1$ if a document $D$ is judged as relevant to $Q$, $\mathcal{Q}$ is a set of benchmark queries, $\mathbb{I}(.)$ denotes the indicator function, and $R_{max}(\mathcal{Q})$
represents the total number of relevant documents known for a static collection, e.g., the ones obtained by employing $\zeta(Q)=100\, \forall Q \in \mathcal{Q}$, i.e, the true depth used to compile the ground-truth of the TREC Robust topic sets. 

We also measure the \textbf{average pool size} $\overline{|\mathcal{P}|}$, as the number of unique documents occurring in a depth restricted pool - again averaged over the queries. Note that this measure is related to the assessment effort. Formally,
\begin{equation}
\overline{|\mathcal{P}|}=
\frac{1}{\mathcal{Q}}
\sum_{Q \in \mathcal{Q}}
|\bigcup_{i=1}^pM_i^{\zeta(Q)}|
\label{eq:aps},
\end{equation}

Since a high coverage (Equation \ref{eq:recall}) and a low average pool size (Equation \ref{eq:aps}) indicate an effective pooling strategy, for the sake of convenient comparisons we combine these two measures into a single metric.
Since the average pool size per query is much larger than the recall values (bounded in $[0,1]$), we compute the ratio after taking a $\log$ of the average pool size, akin to the tf-idf combination where the document frequencies being much larger than the tfs are used with a $\log$ transformation. Formally, we define Pool-size Normalized Coverage (\textbf{PNC}) as
$\mathcal{C}/\log \overline{|\mathcal{P}|}$, a higher value of which indicates a better coverage obtained with a low average pool size. 

\begin{table}[t]
\centering
\small
\caption{\small
A comparison of QPP-based VDP with CDP approaches (including the oracle case, denoted as `AL' or apex-line, shown in green) on the TREC Robust topic sets.
The best results along each column of the non-oracle results are bold-faced. A higher value of all the metrics except $\overline{|\mathcal{P}|}$ (average pool size) indicates a more effective pooling strategy. 
}
\label{tab:mainres_trec}
\begin{adjustbox}{width=.99\columnwidth}
\begin{tabular}{@{}lll r cc ccc@{}}
\toprule
& & & Avg. &  \multicolumn{5}{c}{Evaluation Metrics} \\
\cmidrule{5-9}
Set & Type & Pool & Depth & P-$r$ & K-$\tau$ & $\mathcal{C}$ & $\overline{|\mathcal{P}|}$ & PNC \\
\midrule
\multirow{6}{*}{\begin{turn}{90}TREC 7\end{turn}} & \multirow{2}{*}{BL} & CDP-Min & 10	& 0.9897	& 0.9261	& 0.3988 & \textbf{187.66} & 0.0762 \\
& & CDP-Avg & 30	& \textbf{0.9988} &0.9714 &0.6711 & 484.00 & 0.1086 \\
\cmidrule{2-9}
& \multirow{2}{*}{Ours}& VDP-L & 16.36 & 0.9985 & \textbf{0.9760} & \textbf{0.7021} & 579.52 & \textbf{0.1104} \\
& &VDP-IL & 42.72 & 0.9986	& 0.9718	&0.6467	& 414.50 & 0.1073 \\
\cmidrule{2-9}
& AL & CDP-Max & 50	& \cellcolor{green!25}0.9996	&\cellcolor{green!25}0.9886	&\cellcolor{green!25}0.8223 & \cellcolor{green!25} 759.00 & \cellcolor{green!25}0.1240 \\
%
\midrule
\midrule
\multirow{6}{*}{\begin{turn}{90}TREC 8\end{turn}} & \multirow{2}{*}{BL} & CDP-Min & 10	& 0.9922 & 0.9215&0.4052&	\textbf{239.54}&	0.0740 \\
& & CDP-Avg & 30	& 0.9987 &0.9683&	0.6596&	620.96&	0.1026 \\
\cmidrule{2-9}
& \multirow{2}{*}{Ours}& VDP-L & 22.90	&0.9978	&0.9680	&\textbf{0.6827}	&721.00&	0.1037 \\
& &VDP-IL & 36.14 & \textbf{0.9991}	&\textbf{0.9714}	&0.6600&	540.00&	\textbf{0.1049} \\
\cmidrule{2-9}
& AL & CDP-Max & 50	& \cellcolor{green!25}0.9997	&\cellcolor{green!25}0.9864	&\cellcolor{green!25}0.8201 & \cellcolor{green!25}959.62 & \cellcolor{green!25}0.1194 \\
\bottomrule
\end{tabular}
\end{adjustbox}
\end{table}

\begin{table}[t]
\centering
\small
\caption{\small
Evaluation on TREC DL topic sets, the organization of the table being identical to that of Table \ref{tab:mainres_trec}.}
\label{tab:mainres_dl}
\begin{adjustbox}{width=.99\columnwidth}
\begin{tabular}{@{}lll r cc ccc@{}}
\toprule
& & & Avg. &  \multicolumn{5}{c}{Evaluation Metrics} \\
\cmidrule{5-9}
Set & Type & Pool & Depth & P-$r$ & K-$\tau$ & $\mathcal{C}$ & $\overline{|\mathcal{P}|}$ & PNC\\
\midrule
\multirow{6}{*}{\begin{turn}{90}TREC DL'19\end{turn}} & \multirow{2}{*}{BL} & CDP-Min & 1	& 0.9022	& 0.6336	& 0.2229 & \textbf{8.65} & 0.1033\\
& & CDP-Avg & 3	& 0.9559	&0.7147	&0.4703 & 20.46 & 0.1558 \\
\cmidrule{2-9}
& \multirow{2}{*}{Ours}& VDP-L & 3.37 & \textbf{0.9686} & \textbf{0.8559} & \textbf{0.5398} & 24.76 & \textbf{0.1682} \\
& &VDP-IL & 1.67 & 0.9241 & 0.7297&	0.2814 & 10.83 & 0.1181 \\
\cmidrule{2-9}
& AL & CDP-Max & 5	& \cellcolor{green!25}0.9850	&\cellcolor{green!25}0.9399	&\cellcolor{green!25}0.6542 & \cellcolor{green!25} 30.67 & \cellcolor{green!25} 0.1911\\
%
\midrule
\midrule
\multirow{6}{*}{\begin{turn}{90}TREC DL'20\end{turn}} & \multirow{2}{*}{BL} & CDP-Min & 1	& 0.9760	& 0.8656	& 0.2448 & \textbf{12.27} & 0.0976 \\
& & CDP-Avg & 3	& 0.9944&	0.9299 & 0.4878 & 29.48 & 0.1442\\
\cmidrule{2-9}
& \multirow{2}{*}{Ours}& VDP-L & 3.87 & \textbf{0.9960} & \textbf{0.9334} & \textbf{0.5740} & 37.50 & \textbf{0.1584}\\
& &VDP-IL & 1.16 & 0.9866 & 0.9030&	0.3161 & 15.75 & 0.1146\\
\cmidrule{2-9}
& AL & CDP-Max & 5	& \cellcolor{green!25}0.9977	&\cellcolor{green!25}0.9568	&\cellcolor{green!25}0.6671 & \cellcolor{green!25}45.24 & \cellcolor{green!25}0.1750 \\
\bottomrule
\end{tabular}
\end{adjustbox}
\end{table}

\subsection{Results}

Tables \ref{tab:mainres_trec} and \ref{tab:mainres_dl} present the results of our experiments on the TREC Robust and the TREC DL datasets. We observe the following trends in the results.
First, our proposed variable-depth pooling (VDP) approaches outperform the constant depth pooling approaches with the depth being set to minimum and average values of the depth range, as can be seen from the better correlation values measured with $r$ and $\tau$. Moreover, these high correlations are observed with better coverage and PNC, which answers \textbf{RQ-1} in affirmative. CDP-Max, the apex-line setting (shown as the green rows in the tables) yields better results at the cost of higher annotation effort ($\overline{|\mathcal{P}|}$).

Second, in relation to \textbf{RQ-2}, it can be observed that there is no clear winner between the VDP-L (Equation \ref{eq:vardepth_direct}) and VDP-IL (Equation \ref{eq:vardepth}) variants. While the linear dependence method works better than its inverse-linear counterpart for 3 topic-sets (TREC 7 and the two TREC DL sets as seen by the higher rank correlation, coverage and PNC), the inverse linear works slightly better for TREC 8.

Third, CDP approaches are more robust when the depth is a relatively high value; this can be seen from the better improvements in the Kendall's $\tau$ rank correlations of systems observed on the TREC DL topic sets in comparison to the TREC Robust ones (compare the CDP-Avg values with the VDP ones). It can thus be concluded that VDP approaches are more suitable in cases where the depth range used for VDP is comprised of smaller values, as is the case for the TREC DL ($[1, 5]$) vs. TREC Robust ($[10, 50]$).

Lastly, the VDP methods consistently yield better values of PNC (Pool-size normalized coverage) in comparison to the baseline CDP approaches. This means that a higher number of relevant documents could be found for a set of benchmark queries with reduced manual assessment effort. 

\para{Concluding Remarks}
In this initial investigation of employing variable depth-pooling (VDP) strategies for constructing ground-truth relevance data, our experiments demonstrated encouraging trends. Specifically, we observe that a standard unsupervised QPP method, such as NQC, leads to satisfactory results in terms of high correlation of relative system ranks and also high coverage at the expense of smaller average pool-size. Thus, this indicates stable evaluation results with minimized annotation effort.

There are a number of ways in which we can extend this initial exploration. First, it would be interesting to compare the relative effects of different QPP methods on VDP. It would also be interesting to see if the use of query variants, such as \cite{query_variants_kurland,query_variants_suchana}, can further optimise the depth prediction of VDP.

\para{Acknowledgement}
The first
author thanks Procheta Sen (University of Liverpool) for suggesting this idea during an informal conversation.

\bibliographystyle{ACM-Reference-Format}
\bibliography{ref}
\end{document}